\documentstyle[psfig,12pt,twocolumn]{article}

\begin{document}

\title{$E = mc^2$. Or Is It? : A Comment on Manuscript ASTRO-PH/0504486}
\renewcommand{\baselinestretch}{0.5}

\author{
{\large Ezzat G. Bakhoum}\\
\\
{\small New Jersey Institute of Technology}\\
{\small P.O. Box 305, Marlton, NJ. 08053 USA}\\
{\small Email: bakhoum@modernphysics.org}\\
\\
{\small Copyright \copyright 2005 by Ezzat G. Bakhoum}\\
\\
\\
\begin{minipage}{6in}
\begin{center}
{\bf Abstract}
\end{center}
{\small 
Manuscript astro-ph/0504486 \cite{Gould} is a clear example of the fundamental conceptual flaw that had persisted for the past 100 years. Namely, the failure to realize that mass-energy equivalence, as applied to a propagating light particle (photon), is only a special case of a more general law that encompasses both photons and subluminal particles.}
\end{minipage}
}

\date{}

\maketitle

This is a comment on manuscript astro-ph/0504486 \cite{Gould}, which presents a very interesting derivation of the famous equation $E=mc^2$. This paper, however, is a clear example of the fundamental conceptual flaw that had persisted for the past 100 years. Namely, the failure to realize that mass-energy equivalence, as applied to a propagating light particle (photon), is only a special case of a more general law that encompasses both photons and subluminal particles. More specifically, the fact that a propagating photon carries the equivalent of a ``mass" and that such equivalence is described by the relationship $E=mc^2$, {\em does not automatically imply} that the relationship is valid for particles of ordinary matter. Now, how about the law of conservation of energy? Without a doubt, the law of conservation of energy is valid. The fundamental misconception, however, is the following: while the quantity $E=mc^2$ describing the energy content in a photon must be conserved, it does not {\em mathematically} take the same form when applied to particles of ordinary matter. In other words, the numerical value of the energy is conserved, {\em not its mathematical expression!} For more in-depth discussion of this fundamental issue, the reader is referred to article \cite{Bakhoum1} by the author, and the subsequent articles \cite{Bakhoum2,Bakhoum6}. As is shown in those articles, the correct mass-energy equivalence expression that is applicable to subluminal particles is $E=mv^2$, where $v$ is the particle's velocity. If the particle emits electromagnetic energy in the form of radiation then the change in its total energy will be given by $\Delta E = m_1 v_1^2 - m_2 v_2^2$. The error in derivations such as the one in \cite{Gould} is the assumption that the mass of the particle changes but not its velocity! Since the velocity of light is the limiting velocity, then for a particle traveling with a velocity $v \approx c$, the change in energy will be given by $\Delta E \approx \Delta m c^2$. The famous equation is hence a {\em special case} that applies when a particle is moving with a velocity close to $c$. \\
\\
Attached to this fundamental issue is another important issue, which is the flawed assumption that elementary (Dirac) particles decay at rest (and of course the implications of this assumption as far as the principle of Lorentz invariance is concerned). As was discussed in the earlier papers by the author, particles such as neutrons, $\pi^\pm$, etc. can decay at rest, since those particles have structure within (very fast moving quarks), and hence energy. This is in agreement with the equation $E=mv^2$ and does not violate Lorentz invariance. On the other hand, muons cannot decay at rest, since muons are Dirac particles. It must be noted that, to date, there has been no single experiment that proves decisively whether or not muons decay at rest. Even Leon Lederman had pointed out that his experiment on parity violation \cite{Lederman} does not prove that muons decay at rest, since muons can become trapped within the internal magnetic fields of the atoms and hence have a localized velocity (even though the energy loss formulae suggest that the muon has come to a stop). However, if an experiment can show beyond any doubt that the muon does indeed decay at rest, then the muon cannot be a Dirac particle. It must have structure!\\
\\
We would like to point out that all of the so-called ``non-relativistic" derivations of $E=mc^2$ start with the theory of electromagnetism (Poincar\'{e} had in fact demonstrated -before 1905- that $E=mc^2$ can be proved starting from Maxwell's equations \cite{Whittaker}).  If $E= mc^2$ was indeed a universal law of nature, we should be able to prove it without invoking the theory of electromagnetism. I challenge the physics community to show a proof for $E=mc^2$ without invoking the theory of electromagnetism.\\
\\
\\
{\large\bf Acknowledgements:}\\
The author would like to thank Dr. John Field of the University of Geneva for interesting discussions and insights about the topic of particle decays.

\end{document}